\documentclass[twocolumn,showpacs,preprintnumbers,amsmath,amssymb]{revtex4}
\usepackage{graphicx}
\usepackage{dcolumn}
\usepackage{bm}
\begin{document}
\preprint{}
 

\title{Correlations Between Charge Ordering and Local Magnetic Fields in Overdoped YBa$_2$Cu$_3$O$_{6+x}$}

\author{J.E.~Sonier$^{1,3}$, J.H.~Brewer$^{2,3}$, R.F.~Kiefl$^{2,3}$, R.H.~Heffner$^{4}$,
K.~Poon$^{1}$, S.L.~Stubbs$^{5}$, G.D.~Morris$^{4}$, R.I.~Miller$^{2,3}$, W.N.~Hardy$^{2}$,
R.~Liang$^{2}$, D.A.~Bonn$^{2}$, J.S.~Gardner$^{6}$ and N.J.~Curro$^{4}$}
\affiliation{$^1$Department of Physics, Simon Fraser University, Burnaby,
British Columbia V5A 1S6, Canada \\
$^2$Department of Physics and Astronomy, University of British Columbia,
Vancouver, British Columbia V6T 1Z1, Canada \\
$^3$TRIUMF, Vancouver, British Columbia V6T 2A3, Canada \\
$^4$Los Alamos National Laboratory, Los Alamos, New Mexico 87545, USA \\
$^5$Department of Physics, University of Alberta, Edmonton,
Alberta T6G 2J1, Canada \\
$^6$National Research Council, Chalk River, Ontario K0J 1P0, Canada}
\date{\today}
\begin{abstract}
Zero-field muon spin relaxation (ZF-$\mu$SR) measurements were undertaken on
under- and overdoped samples of  superconducting YBa$_2$Cu$_3$O$_{6+x}$ to
determine the origin of the weak static magnetism recently reported
in this system. The temperature dependence of the muon spin relaxation rate
in overdoped crystals displays an unusual behavior in the superconducting
state. A comparison to the results of NQR and lattice structure experiments on 
highly doped samples provides compelling evidence for strong coupling of 
charge, spin and structural inhomogeneities.
\end{abstract}
\maketitle

There is growing evidence that phase segregation of doped charges
may be an intrinsic property of high-$T_c$ superconductors. 
While the static spin/charge stripes \cite{Stripes}
observed in the marginal superconductor 
La$_{1.48}$Nd$_{0.4}$Sr$_{0.12}$CuO$_4$ \cite{Tranquada:95} 
are expected to be incompatible with the metallic behavior of the 
CuO$_2$ layers \cite{Kivelson:98}, there is evidence from
neutron scattering measurements that a stripe phase that is short range and 
dynamic may exist in several high-$T_c$ systems \cite{McQueeney:99,Neutrons,Petrov:00}.
At low temperatures $\mu$SR experiments indicate a freezing of the Cu spins   
\cite{Weidinger:89,Kiefl:89,Niedermayer:98,Panagopoulos:00}. Local static 
magnetic order observed in highly underdoped samples is 
gradually destroyed with increased hole concentration, giving rise to a spin-glass-like state. 
Niedermayer {\it et al.} found that the transition temperature for spin freezing in the 
La$_{2-x}$Sr$_x$CuO$_4$ and Y$_{1-x}$Ca$_x$Ba$_2$Cu$_3$O$_6$ 
systems shows a similar reduction with increased hole doping \cite{Niedermayer:98}. 
Above the spin freezing temperature Panagopoulos {\it et al.} observed slow spin 
fluctuations that persist slightly above optimal doping \cite{Panagopoulos:00}.
Recently, Pan {\it et al.} observed nano-scale spatial variations 
in the electronic state at the surface of optimally doped 
Bi$_2$Sr$_2$CaCu$_2$O$_{8+\delta}$ using scanning tunnelling microscopy 
(STM) \cite{Pan:01}. Together these observations are consistent with nano-scale 
phase segregation into antiferromagnetic (AF) hole-poor clusters surrounded 
by nonmagnetic hole-rich regions. 

In the YBa$_2$Cu$_3$O$_{6+x}$ (Y123) system there is the added complication
of charge ordering in the CuO chains. Nuclear quadrupole resonance (NQR) 
\cite{Kramer:99,GrevinA:00,Kramer:00,GrevinB:00}, STM \cite{Edwards:95} and 
neutron \cite{Mook:96} measurements on {\it highly doped} 
Y123 are compatible with the formation of a charge density wave (CDW) state 
in the CuO chains. The NQR studies provide strong evidence that the CDW chain state 
induces a charge density modulation in the CuO$_2$ planes \cite{GrevinB:00}.

While early ZF-$\mu$SR studies found no evidence 
for electronic moments in Y123 for $x \! \geq \! 6.54$ \cite{Kiefl:89,Kiefl:90},
recently we detected weak static magnetism of unknown origin in 
$x \! = \!0.67$ and $x\! = \! 0.95$ single crystals \cite{Sonier:01}. While these
findings were discussed primarily in the context of orbital current models 
for the pseudogap phase \cite{Pseudogap}, we noted that our findings may be 
characteristic of a dilute spin system, and thus incompatible with
such theories. Here we present new measurements of under- and overdoped 
Y123 crystals using an improved experimental arrangement. New features that
are clearly observed in the overdoped sample provide strong evidence that charge
ordering is accompanied by the formation of local magnetism, and that these
effects are related to the anomalous local structural changes that have been
observed in this system.

Small single crystals of YBa$_2$Cu$_3$O$_{6+x}$ [$x \! = \! 0.80$, 0.92, 0.985 with
$T_c \! = \! 85 (1.2)$~K, 93.0(0.3)~K and 90.2 (0.8)~K, respectively]
$\sim 100$~$\mu$m thick were grown in BaZrO$_3$ crucibles.
The $x \! = \! 0.80$ and $x \! = \! 0.985$ crystals were mechanically detwinned.
The $\mu$SR  measurements were carried out on the M20B surface muon beam line at
TRIUMF with the initial muon spin polarization {\bf P}$_{\mu}(0)$ perpendicular to
the $\hat{c}$-axis of the crystals. The LAMPF spectrometer was used with a side-axis
low background insert (rather than the axial configuration of Ref.~\cite{Sonier:01}),
resulting in a marked improvement in the quality of the time spectra.
While the $\mu^+$ stopping sites in Y123 have never been firmly established,
we show that correlations with Cu(2) NQR linewidth measurements in
fully doped Y123 imply that the $\mu^+$ is sensitive to magnetism in the
CuO$_2$ planes.

Figure~1 shows the time evolution of the muon spin polarization in overdoped
$x \! = \! 0.985$ crystals in zero external field.
As a spin-1/2 particle, the muon is directly sensitive only to changes
in its magnetic environment.
Thus the increased signal relaxation between 137~K and 10~K indicates
a growth in the size of the local magnetic fields at the $\mu^+$ stopping sites. In addition
to this, we observe a striking change in the shape of the relaxation function near 55~K,
with the signal displaying a slower relaxation at later times. This behavior was not identified 
in our study of lower $x$ samples \cite{Sonier:01}. We find that a longitudinal field (LF) 
of 100~Oe is sufficient to completely decouple the muon spin from the local internal field 
distribution over the entire temperature range. This implies that the magnetism sensed by the 
$\mu^+$ fluctuates at a rate slower than 10$^7$~MHz. 

To identify gross features the ZF time spectra can be fit to a ``stretched exponential'' 
relaxation function $G_z(t) \! = \! \exp[(- \Lambda t)^p]$. Figures~2 and 3 show the 
temperature dependence of the relaxation rate $\Lambda$ and the power $p$, respectively.
Included in these figures are similar fits to the $x \! = \! 0.67$ and $x \! = \! 0.95$ data
of Ref.~\cite{Sonier:01}. The $x \! = \! 0.95$ sample was prepared in
yttria-stabilized zirconia crucibles and was not detwinned. For $x \! = \! 0.985$,
$\Lambda$ and $p$ display an anomalous minimum near 55~K.
Below 35~K, the increase of $\Lambda$ with decreasing $T$ indicates an
increase in the width of the local field distribution. These features become less pronounced 
with decreasing $x$, and are not observed for $x \! = \! 0.67$. The absence of
an obvious minimum in $p$ near 55~K for $x \! = \! 0.95$ reflects the reduced accuracy of
our original measurements.

The relaxation of the muon spin by the randomly oriented
weak dipolar fields of the host nuclei 
is properly described by a static Gaussian Kubo-Toyabe (KT) function
\begin{equation}
G_z^{\rm KT} = \frac{1}{3} + \frac{2}{3}(1-\Delta^2 t^2) \exp ( -\frac{1}{2} \Delta^2 t^2)
\label{eq:KT}
\end{equation}
where $\Delta$ is the width of the field distribution. For $\Delta t \! \ll \! 1$ (which is the
case here) this reduces to $G_z^{\rm KT} \! \approx \! \exp [ (- \Delta t)^2]$. In Fig.~3, 
$p \! \approx \! 2$ at high temperature, but decreases to a smaller value below the 55~K dip. 
This indicates the presence of more than one muon spin relaxation rate. As in 
Ref.~\cite{Sonier:01} we find that the ZF-$\mu$SR time spectra are well fit to the product
\begin{equation}
G_z(t) = G_z^{\rm KT} \exp(- \lambda t).
\label{eq:product}
\end{equation}
For the case of static fields, an exponential relaxation is expected from a dilute random 
distribution of magnetic moments. In Ref.~\cite{Sonier:01} we made
the reasonable assumption that  $\Delta$ was independent of $T$. However, 
to obtain good fits to data in the vicinity of 55~K, it was
necessary to allow $\Delta$ to vary freely with temperature. 
We note that from a careful analysis of ZF-$\mu$SR spectra at high $T$, 
we have determined that the muon ``hops'' at temperatures above 175~K. Previously, 
muon diffusion was clearly identified in Y123 above 200~K \cite{Nishida:90}.
Because a moving $\mu^+$ averages over the fields it sees during its lifetime,
our study is restricted to temperatures below 175~K.

Figure~4 shows the temperature dependence of $\Delta$ and $\lambda$ in the overdoped
sample. Below $\sim \! 130$~K, $\lambda$ increases with decreasing $T$. 
This behavior appears to coincide with the gradual increase of the Cu(2) NQR linewidth
$\delta^{63}\nu_Q(2)$ [see Fig.~4(a)] observed by Gr\'{e}vin {\it et al.} in a fully-doped
($x \! = \! 1.0$) Y123 powder \cite{GrevinA:00,GrevinB:00}. In Ref.~\cite{GrevinB:00} 
the increase in $\delta^{63}\nu_Q(2)$ was attributed to the formation of charge correlations 
in the CuO$_2$ planes that are induced by a CDW transition in the CuO chains. 
A similar finding was reported in an NQR study of  PrBa$_2$Cu$_3$O$_7$ \cite{Grevin:99}. 
The finite value of $\lambda$ below $\sim \! 130$~K indicates that charge ordering 
is accompanied by the onset of local magnetic fields. Because charge ordering is expected to
give rise to strong local Cu spin correlations in the hole depleted regions of the
sample (due to the tendency toward AF order in the underdoped system) we 
hypothesize that the fields are associated with these Cu moments.  

Between 75~K and 175~K, $\Delta$ is independent of $T$. Fits of 
the time spectra to a {\it dynamic} Gaussian KT function 
$G_z^{\rm dKT}$ \cite{Schenck} in place of 
the static function $G_z^{\rm KT}$ of Eq.~(\ref{eq:product}), yielded
$\nu \! \approx \! 0$ at all temperatures below 175~K, where $\nu$ reflects
time-dependent local fields experienced by the $\mu^+$. 
Thus the change in $\Delta$ beginning below 75~K [see Fig.~4(b)] 
is not due to a dynamical relaxation mechanism --- a conclusion 
which is supported by the LF measurements.
The minimum at 55~K and the increase in  $\Delta$ below 35~K  
coincide with features observed in the temperature dependence of $\delta^{63}\nu_Q(2)$ 
[see Fig.~4(a)]. The Cu(2) NQR linewidth displays a maximum near 
60~K, \cite{GrevinA:00,GrevinB:00}, a minimum at 35~K and increases 
monotonically below 35~K with decreasing temperature 
\cite{Kramer:99,GrevinA:00,Kramer:00,GrevinB:00}. In slightly underdoped Y123 
the peak in $\delta^{63}\nu_Q(2)$ near 60~K is not observed  \cite{Kramer:99,Kramer:00}.
The weakening of the 55~K feature in the $\mu$SR data with decreasing $x$
[see Figs.~2 and 3] may be due to an emptying of the CuO chains and/or 
a decreased coupling with the CuO$_2$ planes. 

Gr\'{e}vin {\it et al.} \cite{GrevinB:00} argued that the 60~K peak in 
$\delta^{63}\nu_Q(2)$ arises from a transition between short- and long-range 
charge ordering. The transition is accompanied
by electric field gradient (EFG) and/or magnetic field fluctuations perpendicular 
to the CuO$_2$ planes --- as evidenced by a broad peak in the transverse
relaxation rate $1/T_2$ of the Cu(2) nuclei near 
55~K \cite{Itoh:90,Eremin:01} and also at 35~K \cite{Kramer:99,Itoh:90,Eremin:01,Peak:35}.
Local EFG fluctuations may arise from doped charges that are dynamic.
Although, the $\mu^+$ is sensitive to the effects of EFG fluctuations on the nuclear 
dipole fields, the absence of dynamical relaxation of the muon spin 
polarization is understandable given that the $1/T_2$ maxima measured with NQR are 
about an order of magnitude longer than the $\mu$SR time scale.
High-resolution dilatometry \cite{Meingast:91}, X-ray scattering \cite{You:91}
and microwave \cite{Zhai:01} measurements suggest that the redistribution of charge
may be intimately related to local lattice distortions. 
Below $\sim \! 60$~K (75~K) the $\hat{b}$-axis increases with decreasing $T$  \cite{Meingast:91,You:91}, 
whereas the $\hat{a}$-axis expands below 35~K \cite{You:91}. One interpretation of these 
structural anomalies is that they correspond to an unbuckling of the CuO$_2$ layers \cite{You:91}. 
It is noteworthy that the microwave measurements of Ref.~\cite{Zhai:01} were performed 
on the parent compound YBa$_2$Cu$_3$O$_{6.0}$. This suggests that in highly doped samples 
the lattice distortion taking place near 60~K induces the redistribution of holes.

The unusual $T$ dependence of $\Delta$ below 75~K indicates a change in
the distribution of Cu nuclear dipole moments sensed by the $\mu^+$.
In general, the second moment of this distribution will depend on the relative
angles between {\bf P}$_{\mu}(0)$, the direction of the
EFG at the nuclear site and the orientation of the position vector connecting the nucleus
and the $\mu^+$, as well as the distances between the $\mu^+$ and the host nuclei
\cite{Schenck}. The small change in the tilt angle of the
CuO$_5$ pyramid that accompanies the unbending of the Cu-O-Cu bonds
does not produce a large enough modification of the nuclear dipole
interaction with the $\mu^+$ to account for the observed behavior.
Futhermore, the change in the EFG distribution at the Cu sites
due to the redistribution of charge is too small to produce such a large
change in $\Delta$. A possible explanation for the $T$
dependence of $\Delta$ is that it reflects a change in the distance between the $\mu^+$
and Cu nuclear sites due to Coulomb repulsion between the $\mu^+$ and 
the positive charged holes at the Cu(2) plane site. A variable $\Delta$ 
confounds the quantitative evaluation of the ``extra'' magnetism 
characterized by $\lambda$. Although $\lambda$ is certainly nonzero, 
the increase in relaxation below 35~K previously reported for $x \! = \! 0.95$
\cite{Sonier:01} was at least partly due to an increase in $\Delta$. 
We have now established that the magnetism originates from dilute (quasi) static
electronic moments and not from a dense pattern of orbital currents
spontaneously forming below the pseudogap crossover line $T^*(x)$.
While, some theories \cite{Andergassen:01} suggest a close connection between
the onset of charge ordering and $T^*(x)$, we cannot assign an
unambiguous ``onset'' temperature $T^*$ due to the rich
relaxation phenomenology and the gradual decrease of $\lambda$ with
increasing temperature.

From a comparison with NQR measurements on fully doped Y123,
we have determined that the growth of weak local magnetism
with decreasing $T$ coincides with the evolution of spatial charge 
inhomogeneities. The agreement between $\mu$SR,
NQR and lattice structure experiments implies that this spatial
inhomogeneity is an intrinsic property of Y123 that arises 
independent of the sample preparation details.
Finally, we make the interesting observation that the
anomaly near 60~K appears absent for $x \! = \! 0.67$, despite its
close proximity to the well known ``60~K plateau'' in $T_c$.

This work was supported by the Natural Sciences and Engineering Research
Council of Canada, the Canadian Institute for Advanced Research,
and at Los Alamos by the US Department of Energy. We thank
W.A.~MacFarlane and Z.~Yamani for fruitful discussions.
We are especially grateful to Syd Kreitzman, Mel Good, Donald
Arseneau and Bassam Hitti for technical assistance.

\newpage
\begin{center}
FIGURE CAPTIONS
\end{center}

Figure 1. The time evolution of the muon spin polarization in
YBa$_2$Cu$_3$O$_{6.985}$ in zero external field at $T \! = \! 10$,
56 and 137~K. The signal at 10~K relaxes faster than at 137~K.
At 56~K the signal at early times relaxes faster than at 137~K, but
exhibits a slower relaxation beyond 5~$\mu$s.\\

Figure 2. Temperature dependence of the relaxation rate $\Lambda$.
Open circles are from fits to the data of Ref.~\cite{Sonier:01}.\\

Figure 3. Temperature dependence of the power $p$.
Open circles are from fits to the data of Ref.~\cite{Sonier:01}.\\

Figure 4. Temperature dependence of (a) $\delta^{63}\nu_Q(2)$ in 
an $x \! = \! 1.0$ powdered sample (from Ref.~\cite{GrevinB:00}) 
and of (b) $\Delta$ and (c) $\lambda$ from fits of the ZF-$\mu$SR 
time spectra in $x \! = \! 0.985$ crystals to Eq.~(\ref{eq:product}).\\


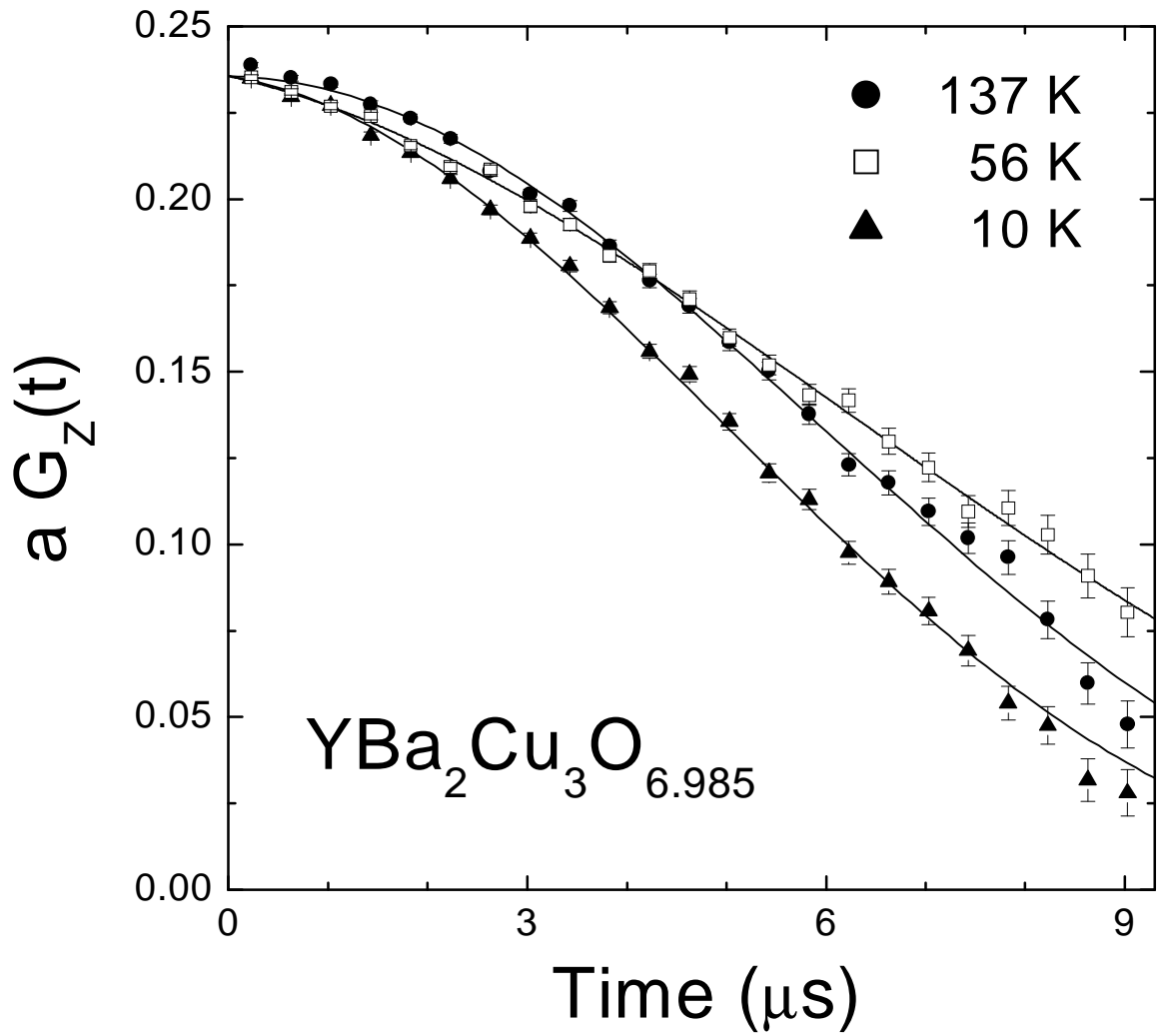
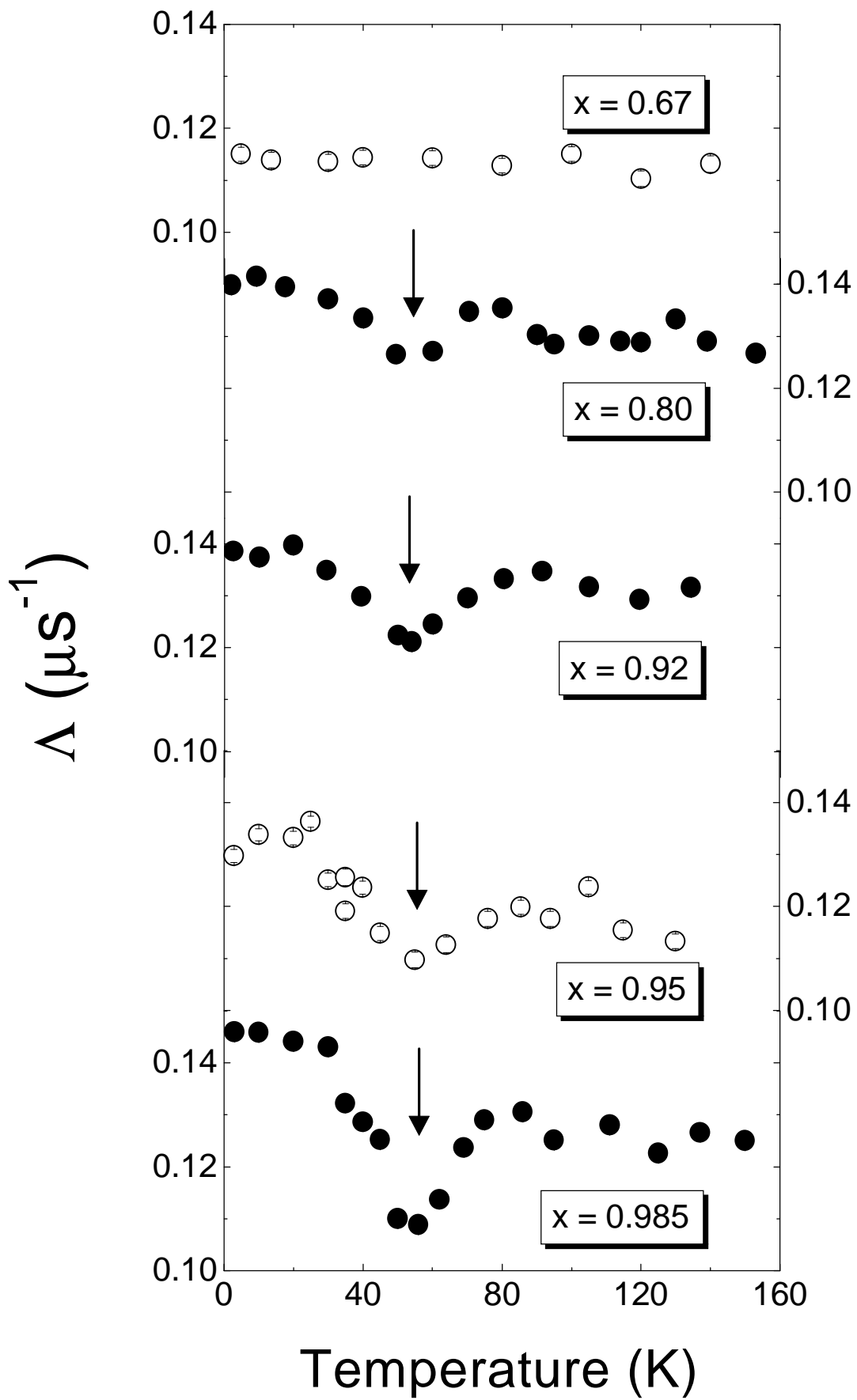
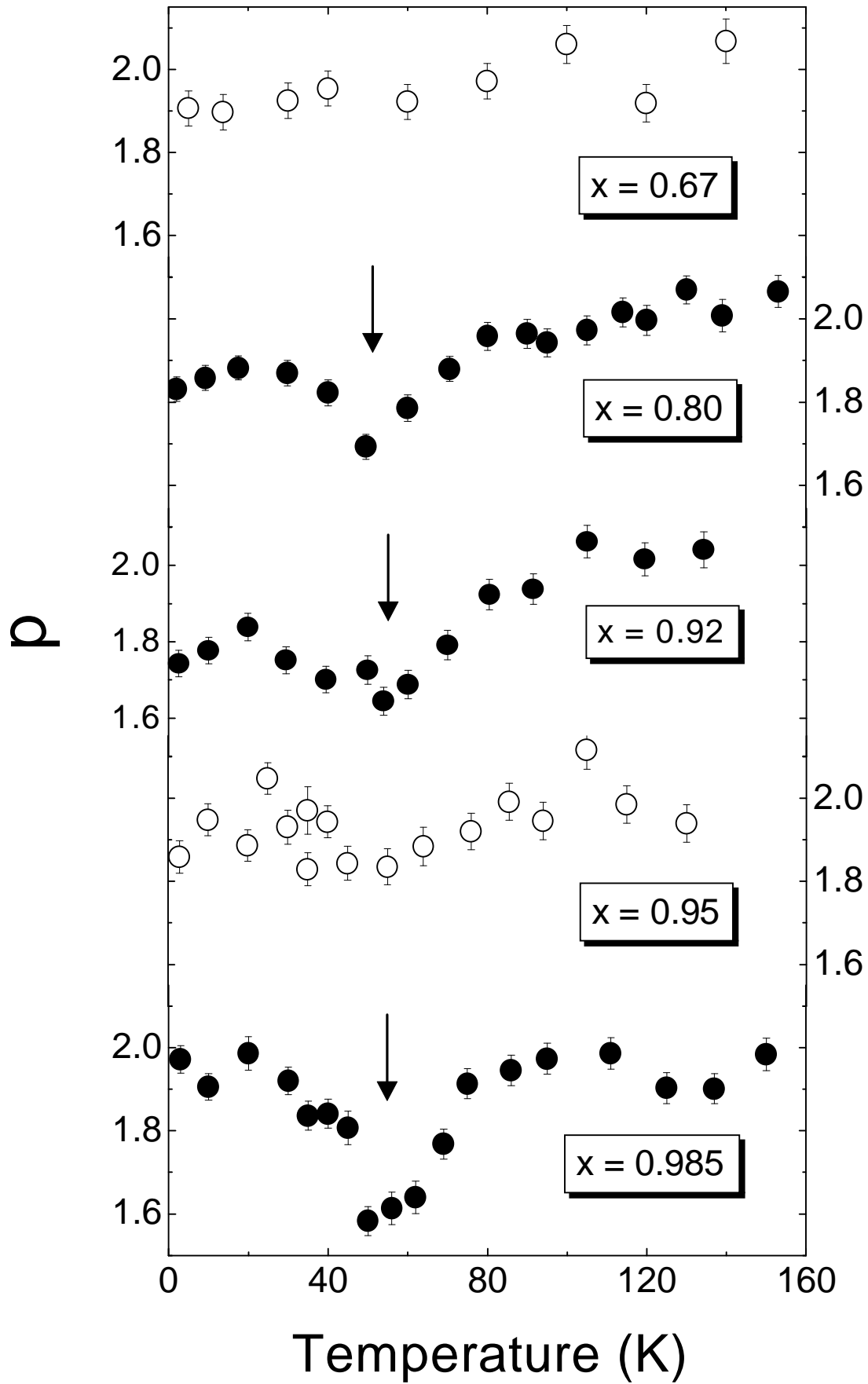
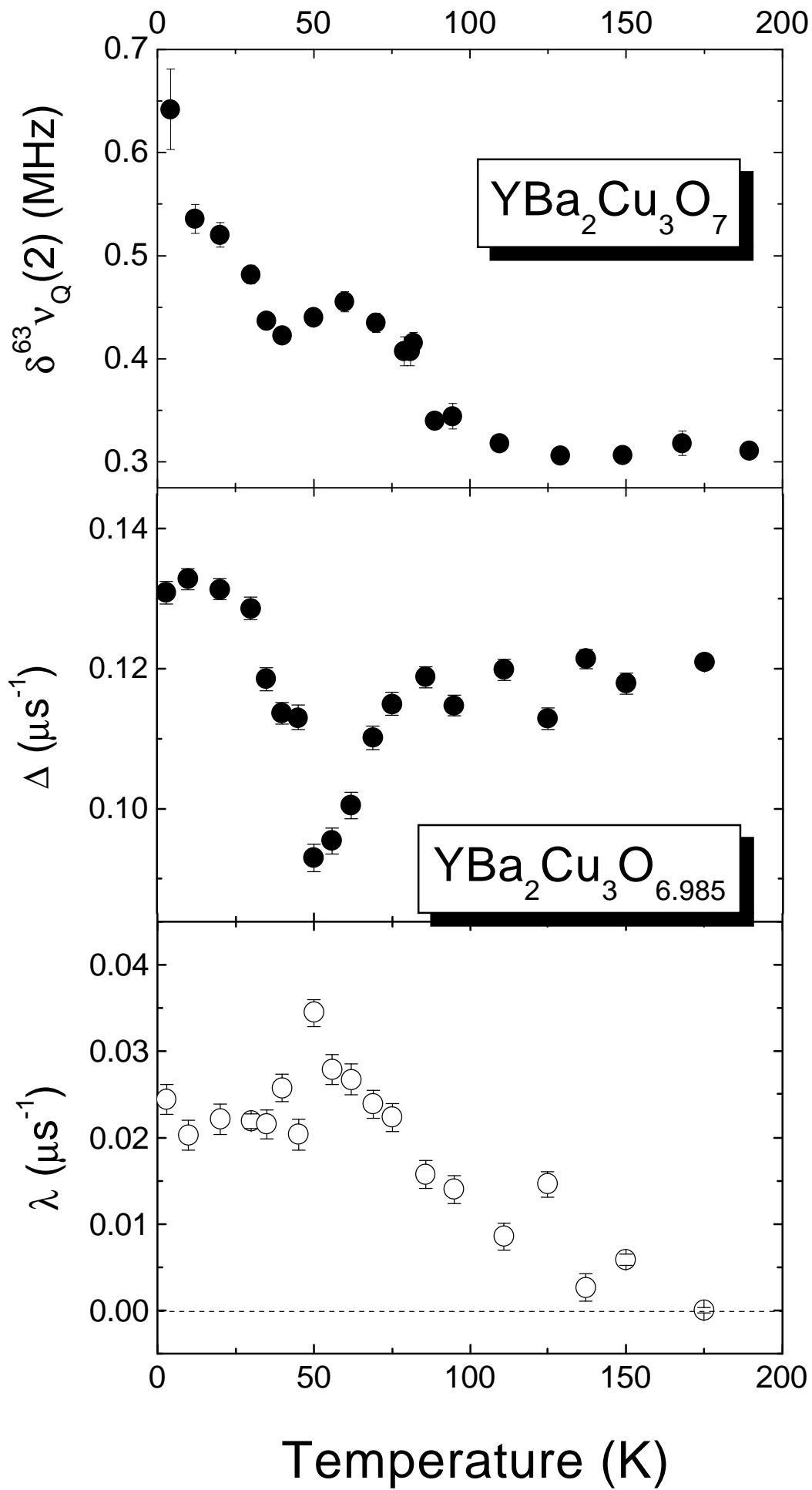
\end{document}